\documentclass[submission, Proceedings]{SciPost}

\begin{document}

\begin{center}{\Large \textbf{
Tau Neutrinos in IceCube, KM3NeT and the Pierre Auger Observatory
}}\end{center}

\begin{center}
D. van Eijk\textsuperscript{1}
\end{center}

\begin{center}
{\bf 1}  Wisconsin IceCube Particle Astrophysics Center, University of Wisconsin, Madison, WI 53703, USA
\\
* daan.vaneijk@icecube.wisc.edu
\end{center}

\begin{center}
\today
\end{center}

\definecolor{palegray}{gray}{0.95}
\begin{center}
\colorbox{palegray}{
  \begin{tabular}{rr}
  \begin{minipage}{0.05\textwidth}
    \includegraphics[width=8mm]{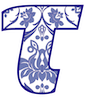}
  \end{minipage}
  &
  \begin{minipage}{0.82\textwidth}
    \begin{center}
    {\it Proceedings for the 15th International Workshop on Tau Lepton Physics,}\\
    {\it Amsterdam, The Netherlands, 24-28 September 2018} \\
    \href{https://scipost.org/SciPostPhysProc.1}{\small \sf scipost.org/SciPostPhysProc.Tau2018}\\
    \end{center}
  \end{minipage}
\end{tabular}
}
\end{center}


\section*{Abstract}
{\bf
In 2018, the IceCube collaboration reported evidence for the identification of a blazar as an astrophysical neutrino source. That evidence is briefly summarised here before focusing on the prospects of tau neutrino physics in IceCube, both at high energies (astrophysical neutrinos) and at lower energies (atmospheric neutrino oscillations). In addition, future neutrino detectors such as KM3NeT and the IceCube Upgrade and their tau neutrino physics potential are discussed. Finally, the detection mechanism for high-energy (tau) neutrinos in the Pierre Auger Observatory and the resulting flux upper limits are presented.}

\vspace{10pt}
\noindent\rule{\textwidth}{1pt}
\tableofcontents\thispagestyle{fancy}
\noindent\rule{\textwidth}{1pt}
\vspace{10pt}

\section{Introduction}
\label{sec:intro}
The discovery of an astrophysical neutrino flux in 2013 by the IceCube collaboration marked the birth of neutrino astronomy. Ever since, both the ANTARES and IceCube collaborations have been searching for the sources of these astrophysical neutrinos. This endeavour paid off in 2018, when IceCube gathered evidence for a known blazar as an astrophysical neutrino source. This result was obtained in close collaboration with other astronomical observatories, which shows both the maturity and the power of the multi-messenger approach to astronomy. 

In addition to these breakthrough discoveries, IceCube currently has a rich tau neutrino physics program. At the same time, tau neutrino physics is one of the main drivers for future neutrino detectors such as the IceCube Upgrade and KM3NeT. Finally, the Pierre Auger Observatory is sensitive to high energy (tau) neutrinos in an energy range complementary to that of IceCube, ANTARES and KM3NeT ARCA.

Section \ref{sec:detectors} contains a brief overview of the IceCube and KM3NeT detector layouts. Next, the evidence for an astrophysical neutrino source by IceCube is briefly summarised in Section \ref{sec:neutrinosource}. Sections \ref{sec:highetausinicecube} and \ref{sec:lowetausinicecube} cover  tau neutrino physics prospects in IceCube. Section \ref{sec:icuandkm3net} motivates the current construction of KM3NeT and the future construction of the IceCube Upgrade from a tau neutrino physics perspective. Finally, Section \ref{sec:nusinpa} summarises the potential of and latest results for the detection of ultra-high energy (tau) neutrinos using the Pierre Auger Observatory.

\section{The IceCube and KM3NeT Neutrino Detectors}\label{sec:detectors}

\subsection{IceCube}\label{sec:icecubedetector}
The IceCube detector is embedded in the Antarctic ice at the geographical South Pole. It consists of 86 vertical strings that are frozen into boreholes. The strings are arranged in a hexagonal grid with 125 meter spacing. At depths between 1450 and 2450 meters, the vertical strings are instrumented with so-called digital optical modules (DOM), a spherical glass pressure vessel containing a 10-inch downward-looking PMT and all associated electronics. Each string holds 60 DOMs, interspaced at a vertical distance of 17 meters. 

\begin{figure}[htb]
\centering
\includegraphics[width=0.8\textwidth]{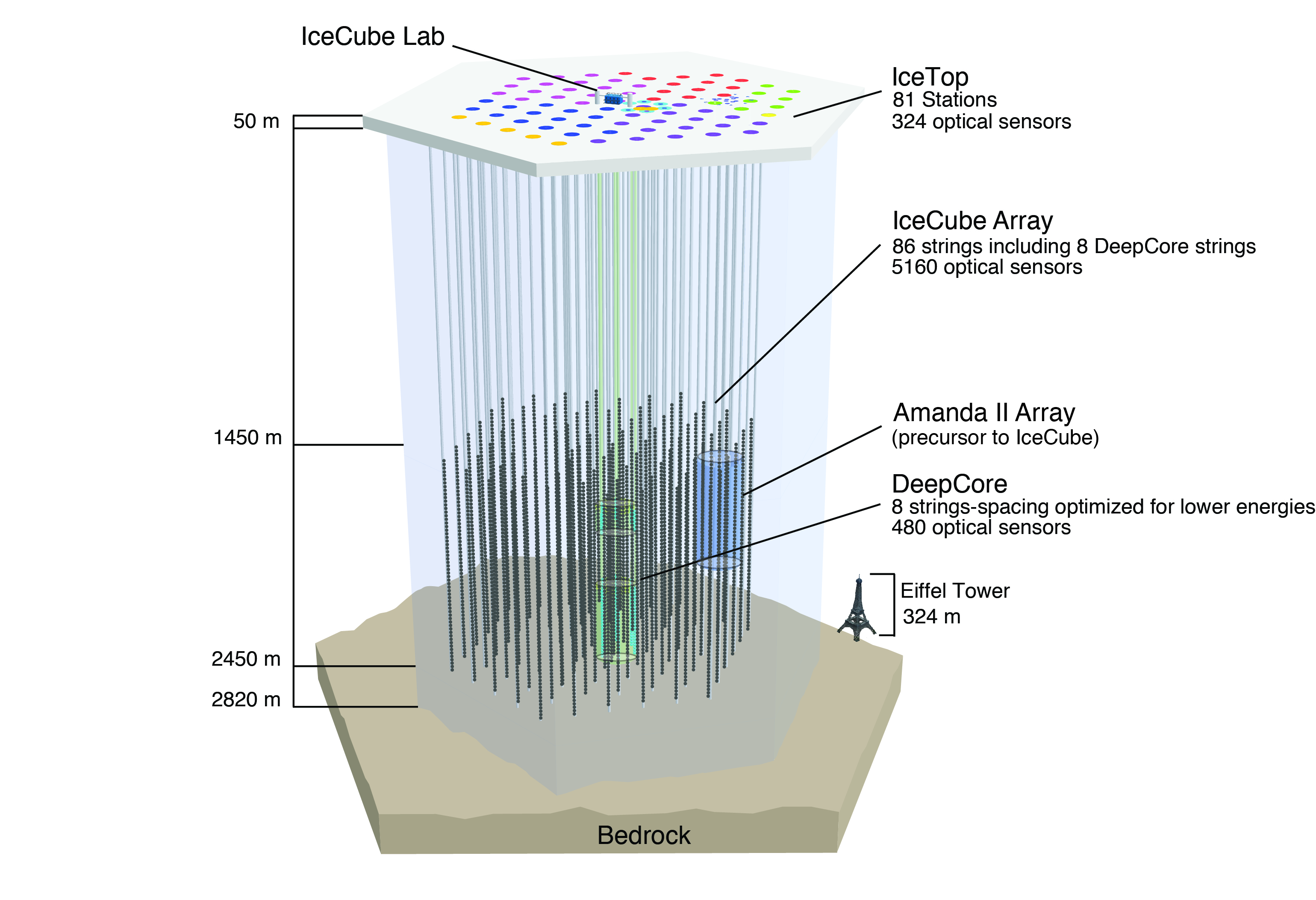}
\caption{Schematic picture of the IceCube detector, showing the position of Amanda, IceCube's predecessor, the dense infill array IceCube DeepCore and the IceTop surface detector for the detection of cosmic rays. All data is collected in the IceCube Lab on the ice surface.}
\label{fig:icecube}
\end{figure}

In the center of the IceCube detector eight strings are deployed closer together at 70 meter distance, containing DOMs at a smaller vertical spacing of 7 meters. This denser infill array, called IceCube DeepCore has a lower neutrino energy detection threshold, allowing for the study of atmospheric neutrino oscillations. A schematic picture of the IceCube detector is shown in Figure \ref{fig:icecube}.

In essence, IceCube is an enormous three dimensional array of 5160 optical modules, allowing for the detection of Cherenkov light that is produced by secondary particles traversing the ice following a neutrino interaction in or around the detector. The arrival time of the Cherenkov light on the various modules is used to reconstruct the direction of the original neutrino, while the amount of light is a proxy for the neutrino's energy. 

\subsection{KM3NeT}\label{sec:km3netdetector}
\begin{figure}[htb]
\centering
\includegraphics[width=0.4\textwidth]{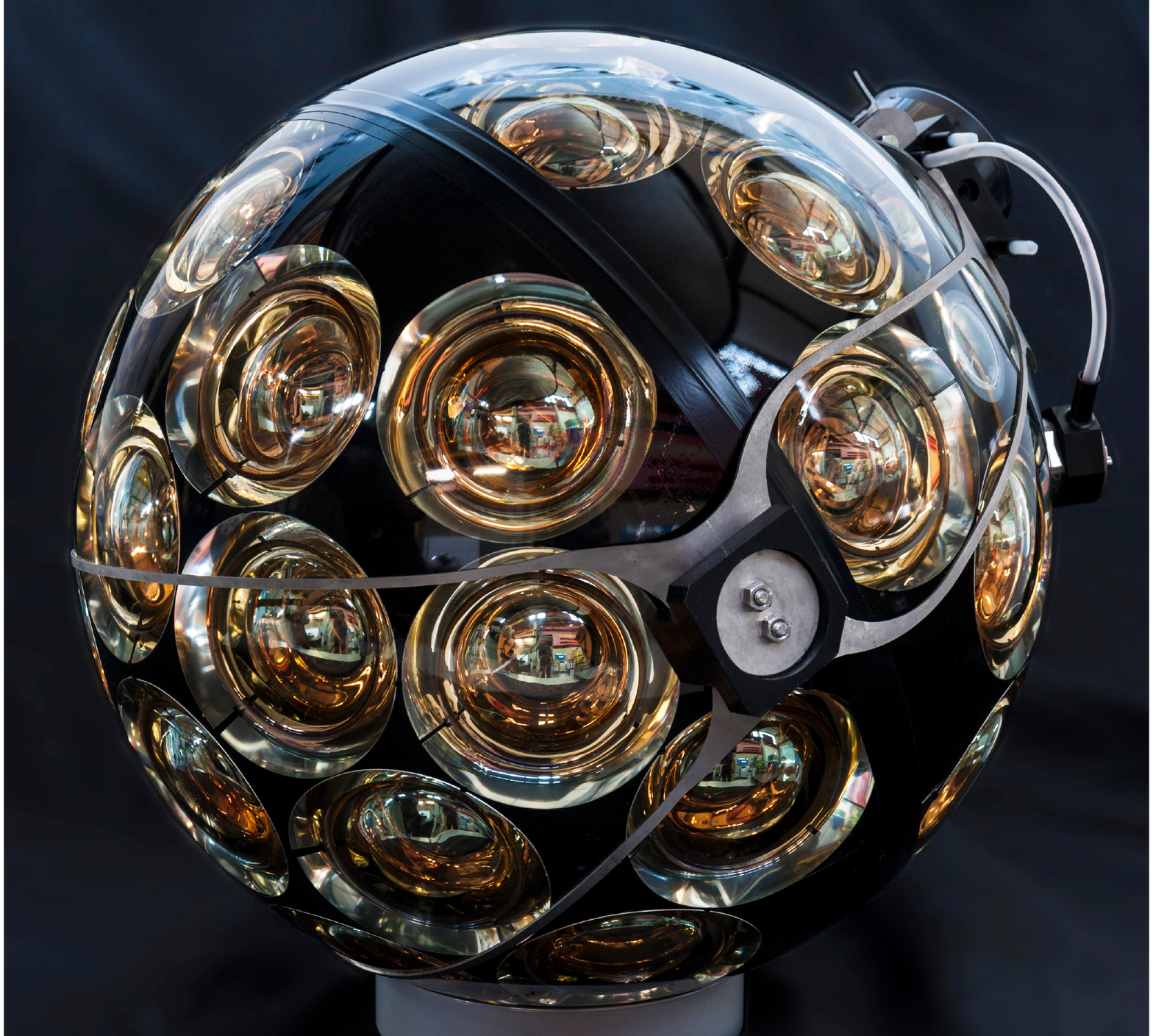}
\includegraphics[width=0.3\textwidth]{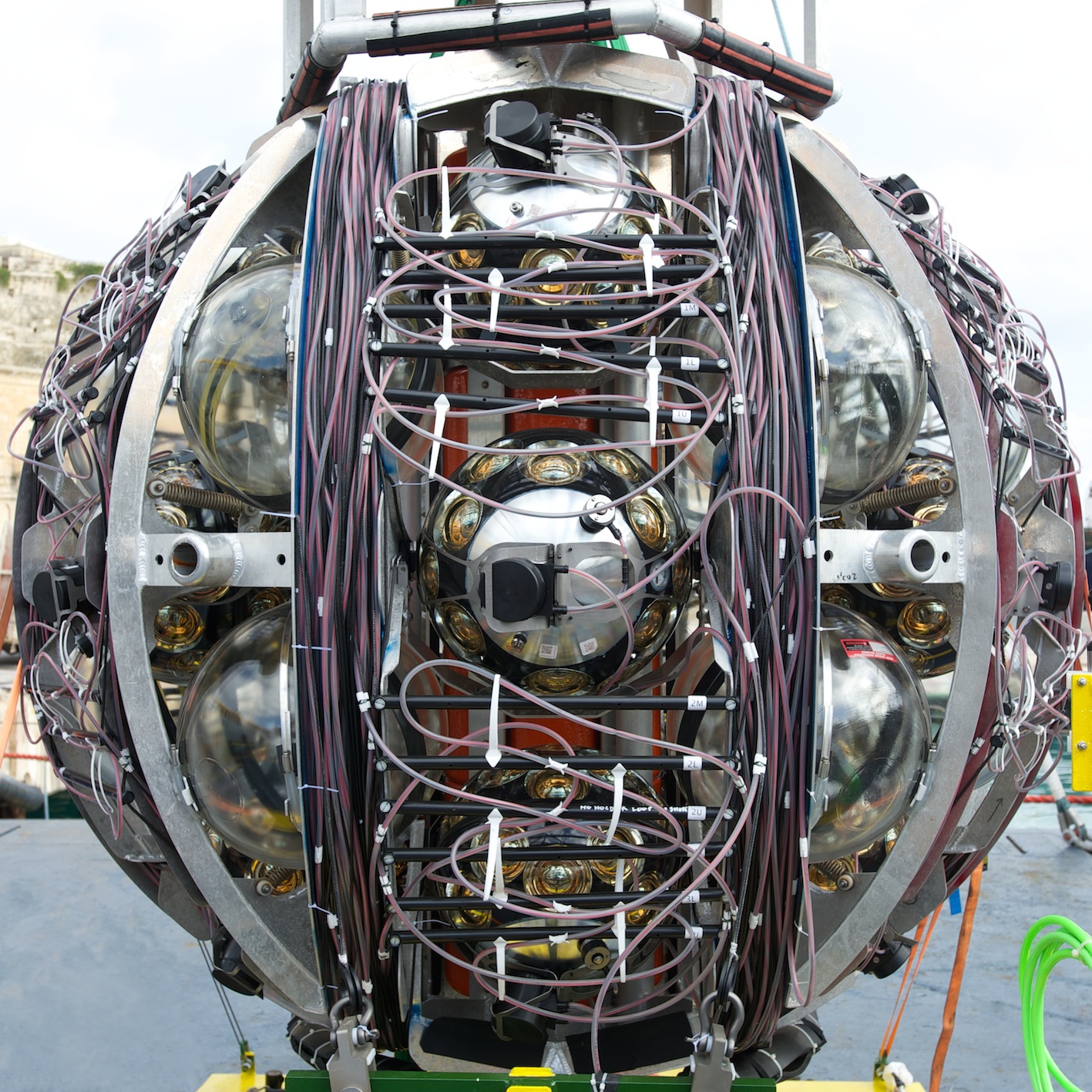}
\caption{Left: a picture of a completed KM3NeT multi-PMT DOM. Right: a fully assembled KM3NeT string ready for deployment, wound up on a deployment vessel.}
\label{fig:km3netdom}
\end{figure}

The general layout of KM3NeT is very similar to that of IceCube: the detector is a large, cubic-kilometre sized three dimensional array of optical modules. KM3NeT is currently under construction at the bottom of the Mediterranean Sea, so the most important difference between IceCube and KM3NeT is the detection medium of water versus ice. Another striking difference is the use of multi-PMT optical modules in the KM3NeT detector \cite{km3netdom}. The KM3NeT DOM contains 31 three-inch PMTs that allow for directional sensitivity, single photon counting and background reduction through PMT multiplicity cuts at the DOM level. A picture of a completed KM3NeT multi-PMT DOM and a full KM3NeT string wound on a deployment vessel is shown in Figure \ref{fig:km3netdom}.

KM3NeT will be constructed at two separate geographical locations: a densely instrumented detector called KM3NeT ORCA will be built off the French coast and will study low-energy atmospheric neutrino oscillations, while a more sparsely instrumented detector called KM3NeT ARCA will be built off the Italian coast near Sicily for the study of high-energy astrophysical neutrinos.

\section{Evidence for an Astrophysical Neutrino Source}\label{sec:neutrinosource}
\subsection{IceCube's Discovery of an Astrophysical Neutrino Flux}\label{sec:astrophysicalflux}
In 2013, IceCube found evidence for an astrophysical neutrino flux \cite{evidenceastrophysicalflux,observationastrophysicalflux} by looking at high-energy neutrino events starting within the detector volume, to reduce background from atmospheric muons and neutrinos. This groundbreaking result was confirmed by an independent analysis in IceCube that studied so-called through-going muons \cite{throughgoing}. In that analysis, atmospheric backgrounds are reduced by only looking at upward-going muons, i.e. neutrinos coming from the Northern hemisphere. Despite this discovery of an astrophysical neutrino flux, until recently it was not possible to tell conclusively where these astrophysical neutrinos originated from.

\subsection{TXS 0506+056}\label{sec:txs}
\begin{figure}[htb]
\centering
\includegraphics[width=0.8\textwidth, bb=50 50 800 600]{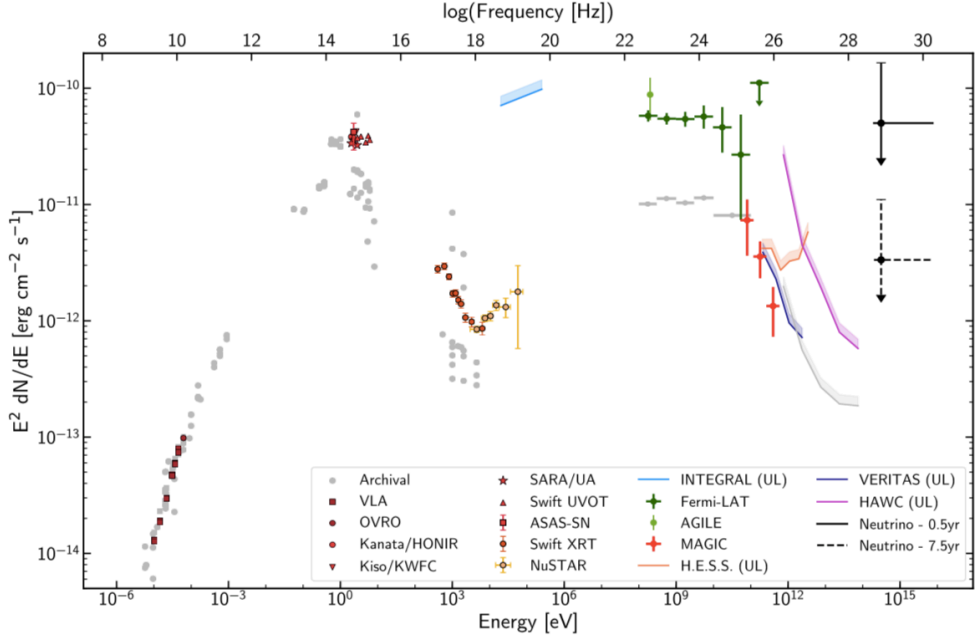}
\caption{Spectral energy distribution for the blazar TXS 0506+056, based on observations obtained within 14 days of IceCube-170922A. Archival observations are shown in gray to illustrate the historical flux level of the blazar in the radio to keV range. Representative $\nu_\mu + \overline{\nu}_\mu$ neutrino flux upper limits that produce on average one detection like IceCube-170922A over a period of 0.5 (solid black line) and 7.5 years (dashed black line) are shown assuming a spectrum of $dN/dE \sim E^{-2}$ at the most probable neutrino energy of 311 TeV. Plot source: \cite{TXSmultimessenger}.}
\label{fig:txssed}
\end{figure}

On September 22 2017, a high-energy neutrino-induced muon track event was detected by IceCube's real-time alert system. This detection, called IceCube-170922A, generated an automated alert that was distributed to the astronomical community to ask for further observations. The alert was later reported to be in positional coincidence with a known $\gamma$-ray blazar TXS 0506+056, that happened to be in a flaring state at the time of the neutrino detection. Combining the extensive multi-wavelength and multi-messenger observations, a chance correlation between this high-energy neutrino and the potential blazar counterpart can be rejected at the 3$\sigma$ level. The resulting multi-messenger spectral energy distribution is shown in Figure \ref{fig:txssed}. More information on the multi-messenger observations can be found in \cite{TXSmultimessenger}. It should be noted that no neutrinos were observed from the direction of IceCube-170922A by the ANTARES neutrino detector (KM3NeT's predecessor) in a $\pm$1 day period around the event.

Based on the temporal and spatial coincidence between IceCube-170922A and TXS 0506+056, IceCube investigated 9.5 years of archived data to search for excess neutrino emission from the blazar's position in the sky. An excess of high-energy neutrino events with respect to atmospheric backgrounds was indeed found between September 2014 and March 2015, constituting to a 3.5$\sigma$ evidence for neutrino emission from the direction of TXS 0506+056. More information on the historical IceCube neutrino flare can be found in \cite{TXSicecubearchive}.

Though this is the first time that evidence has been found for an astrophysical neutrino source, the search for these sources is far from over. First of all, blazars can only partly explain the measured astrophysical neutrino flux \cite{blazarsasflux}. Also, based on this one observed neutrino, the estimated time-integrated neutrino flux from TXS 0506+056 is less than 1\% of IceCube's total observed astrophysical neutrino flux. This means that additional data and additional detectors such as KM3NeT ARCA are much needed. 

\section{High-Energy Tau Neutrinos in IceCube}\label{sec:highetausinicecube}
\subsection{Tau Neutrino Candidates}
There are three distinct neutrino event signatures in large water or ice neutrino detectors. The first is a muon track that results from a charged current interaction of a muon neutrino. The second is a so-called cascade or shower that results after either a neutral current interaction of any neutrino type, or an electron or low energy tau neutrino undergoing a charged current interaction. The third event signature  occurs when a high-energy tau neutrino undergoes a charged current interaction, creating a primary shower. However, the produced tau lepton has a decay length that scales roughly as 1 PeV/50m, causing the formation of a secondary shower when the tau lepton decays to other hadrons and electrons. This event signature is called a 'double cascade', after the two showers separated by the tau decay distance. 

\begin{figure}[htb]
\centering
\includegraphics[width=0.8\textwidth, bb=50 50 800 600]{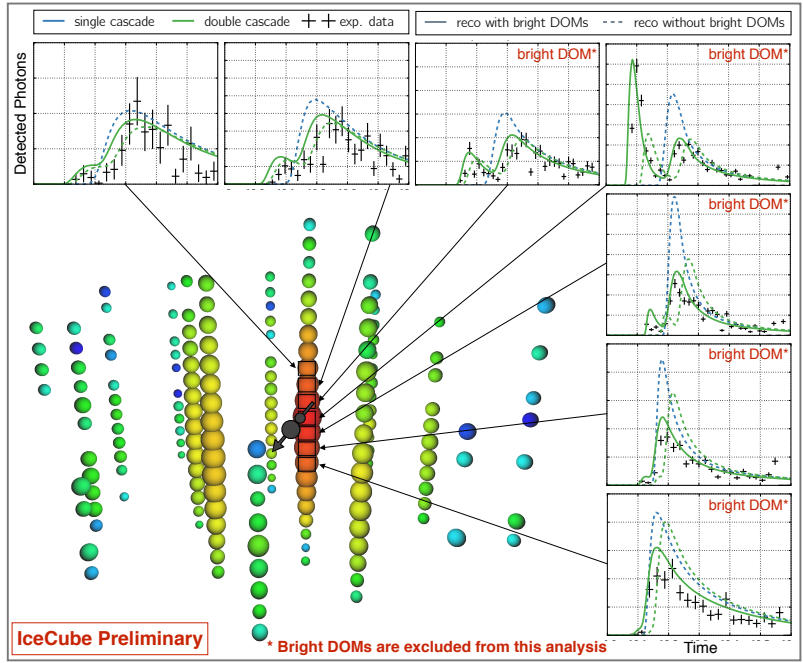}
\caption{Double cascade event display. The colored circles indicate optical modules, where the color scale runs from red (earliest photon arrival time) to green (later photon arrival time). The size of the circle indicates the amount of deposited charge in the optical module. The reconstructed (double) cascade positions are indicated by grey circles and the direction is indicated by a grey arrow. The sizes of the grey circles illustrate the relative deposited energy in the two cascades. The surrounding plots are time distributions of detected photons in selected optical modules in the event. In various optical modules, a double-peaked structure is observed, clearly favouring a double cascade hypothesis. Image source: \cite{HESE75}.}
\label{fig:taucandidate}
\end{figure}

IceCube has performed a search for these double cascade signatures in 7.5 years of high-energy starting events \cite{HESE75} and found two events satisfying the cuts on the reconstructed length, energy asymmetry and energy confinement between the two showers. The expected number of events in the  dataset was 2.1 (1.4 $\nu_\tau$-induced double cascade events and 0.7 background events, from all other decay types). One of the two selected events is shown in Figure \ref{fig:taucandidate}, where the observed light arrival time pattern clearly favours the double cascade hypothesis. These two tau neutrino candidates are further evaluated for possible backgrounds and additional double pulse analyses are ongoing to conclusively identify tau neutrinos in IceCube.

\subsection{Neutrino Oscillations at Cosmic Baselines}
\begin{figure}[htb]
\centering
\includegraphics[width=0.8\textwidth, bb=70 70 800 600]{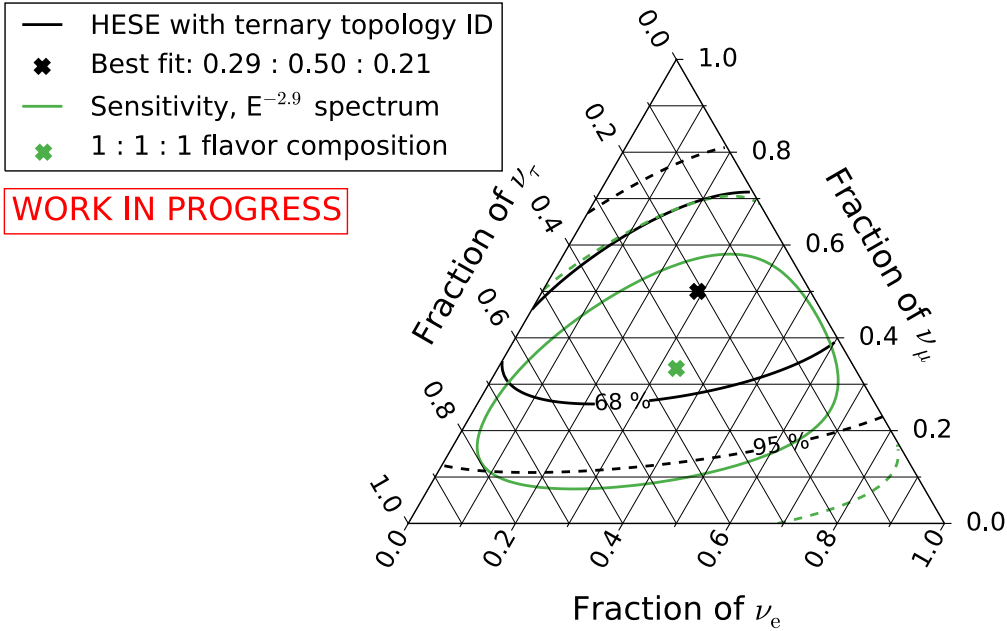}
\caption{Measured flavor composition for 7.5 years of IceCube high-energy starting events with ternary topology ID (i.e. including the double cascade event signature) in black and expected sensitivity at the best-fit spectral index in green. Within the uncertainties, the fit result is consistent with $\nu_{e} : \nu_\mu : \nu_\tau = 1 : 1 : 1$ but a zero $\nu_\tau$ flux can't be excluded either. Plot source: \cite{vlvnt-hese}.}
\label{fig:flavorratio}
\end{figure}

Taking into account neutrino oscillations at cosmic baselines, the expected astrophysical neutrino flavor ratio is expected to be $\nu_{e} : \nu_\mu : \nu_\tau = 1 : 1 : 1$. A fit of the flavor composition of the astrophysical neutrinos in the high-energy starting event dataset, including the double cascade event signature, yields $\nu_{e} : \nu_\mu : \nu_\tau = 0.29 : 0.50 : 0.29$. The corresponding flavor ratio plot is shown in Figure \ref{fig:flavorratio}. The fit result is consistent with $\nu_{e} : \nu_\mu : \nu_\tau = 1 : 1 : 1$ but a zero $\nu_\tau$ flux can't be excluded either.

\section{Low-Energy Tau Neutrinos in IceCube}\label{sec:lowetausinicecube}
Atmospheric neutrino oscillations can be detected using the dense infill array IceCube DeepCore, covering a wide range of neutrino energies and travelled baselines through the Earth, where the cosine of the reconstructed neutrino zenith angle $\theta_z$ is a proxy for the latter. When plotting neutrino oscillation probabilities as a function of neutrino energy (in the range of a few GeV up to 100 GeV) and the distance travelled through the Earth, distinct patterns emerge in narrow regions of this phase space. Taking into account matter effects caused by the Earth's density profile, distortions in the vacuum oscillation probabilities occur for neutrinos travelling straight through the Earth's core, i.e. at $\cos(\theta_z)$ close to -1. These distortions ultimately provide sensitivity to the neutrino mass ordering. Actual IceCube DeepCore data is binned in reconstructed neutrino energy, cosine of the reconstructed zenith angle and PID (track or cascade) to represent the aforementioned oscillograms. From these, standard neutrino oscillation parameters are extracted, with sensitivities comparable to dedicated accelerator and atmospheric neutrino oscillation experiments \cite{icecubeoscillations}. 

\begin{figure}[htb]
\centering
\includegraphics[width=0.8\textwidth, bb=10 10 500 430]{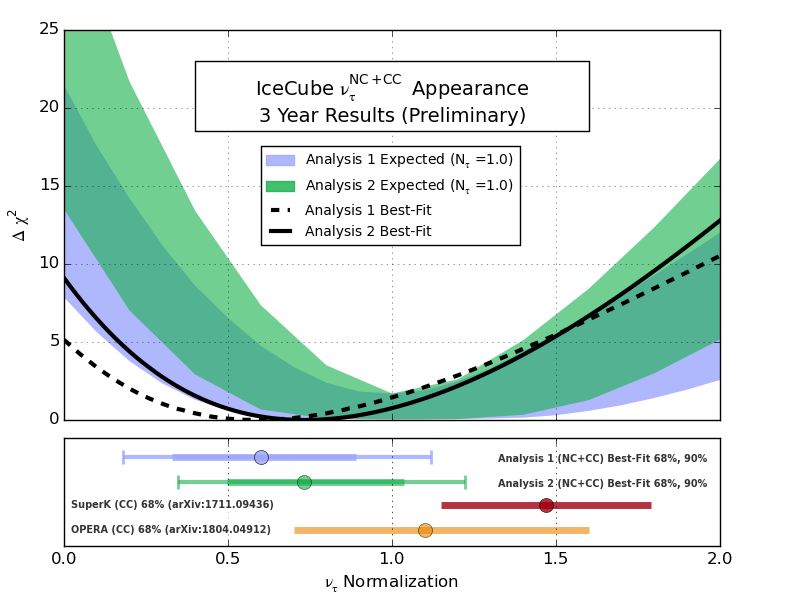}
\caption{Tau normalisation results for two independent IceCube DeepCore analyses (Analysis 1 and Analysis 2) compared to the corresponding OPERA and SuperK results. The Feldman-Cousins method was used in the estimation of the 68\% and 90\% confidence intervals in the bottom part of the plot. Plot source: \cite{vlvnt-oscillations}.}
\label{fig:taunorm}
\end{figure}

The main neutrino oscillation channel studied in IceCube is muon neutrino disappearance. In the standard three-flavor neutrino oscillation paradigm, muon neutrino disappearance is mainly $\nu_\mu \rightarrow \nu_\tau$. At these low neutrino energies, the $\tau$ decays instantly, so $\nu_\tau$ interactions and subsequent tau decays show up in the detector as (single) cascades. By allowing for a $\nu_\tau$ appearance contribution to all cascade events, the so-called tau normalisation can be fitted, which is defined as the ratio of the measured $\nu_\tau$ flux to that expected assuming standard three-flavor neutrino oscillations. The tau normalisation results for two independent IceCube analyses are shown in Figure \ref{fig:taunorm}. 
 
\section{Future Detectors: the IceCube Upgrade and KM3NeT}\label{sec:icuandkm3net}
IceCube is scheduled to deploy seven additional strings within the IceCube DeepCore footprint by 2022-2023. This very densely instrumented detector is called the IceCube Upgrade and predominantly features novel multi-PMT optical modules inspired by the KM3NeT DOM design. In addition, the IceCube Upgrade strings contain improved calibration devices that allow for a better characterisation of the Antarctic ice optical properties. Better knowledge of the absorption and scattering properties of the ice improves the angular resolution for IceCube as a whole, increasing the sensitivity to ongoing emission from point sources of astrophysical neutrinos. The cascade angular resolution is expected to improve from 10$^\circ$-15$^\circ$ to below 5$^\circ$ for neutrino energies above 200 TeV. For KM3NeT ARCA, the corresponding number for the angular resolution is below 2$^\circ$, owing to reduced photon scattering in water compared to ice.

Regarding atmospheric neutrino oscillations, the dense detector spacing of the IceCube Upgrade results in a lower neutrino energy detection threshold compared to IceCube DeepCore, leading to an increased sensitivity to neutrino oscillation parameters, tau normalisation and in particular the determination of the neutrino mass ordering. Table \ref{tab:lowenergysciencegoals} briefly compares selected atmospheric neutrino oscillations science goals for the IceCube Upgrade and KM3NeT ORCA.

\begin{table}[h]
\begin{centering}
\footnotesize
\begin{tabular}{p{5cm}p{3.5cm}p{5cm}}
\hline
\hline
& \textbf{IceCube Upgrade} & \textbf{KM3NeT ORCA}  \\
\hline
Completion & 2022-2023 \cite{vlvnt-icu}	& 2023 \cite{neutrino2018-km3net}\\
Tau normalisation & 13\% constrained & 20\% constrained at 3$\sigma$ after 1 year\\
Neutrino mass ordering (*) & 3$\sigma$ in 3-8 years & 3-6$\sigma$ in 3 years\\
$\Delta m_{32}^2$ relative uncertainty (RU)& 3\% & 3\%\\
$\sin^2(\theta_{23})$ RU at maximal mixing & 14\% & 13\%\\
$\sin^2(\theta_{23})$ RU off-maximal mixing & 8\% & 4\% \\
\end{tabular}
\caption{Table comparing selected low-energy (atmospheric neutrino oscillations) science goals for the IceCube Upgrade and KM3NeT ORCA. All values are at 90\% confidence level after three years of data taking, unless indicated otherwise. (*): depending on the true mass ordering and the true value of $\sin^2(\theta_{23})$.}
\label{tab:lowenergysciencegoals}
\end{centering}
\end{table}


\section{Detecting High-Energy Neutrinos with the Pierre Auger Observatory}\label{sec:nusinpa}
The Pierre Auger Observatory (PAO) in Argentina studies ultra-high energy cosmic rays using a hybrid detector consisting of 27 fluorescence Cherenkov telescopes and 1600 water Cherenkov tanks covering an area of $\sim 3000$\,km$^2$. The existence of cosmic rays at energies above $10^{18}$\,eV implies the existence of an accompanying cosmic neutrino flux that can be created both at the astrophysical source and in interactions with the cosmic microwave background when traveling intergalactic distances.

\begin{figure}[htb]
\centering
\includegraphics[width=0.9\textwidth, bb = 10 10 800 220]{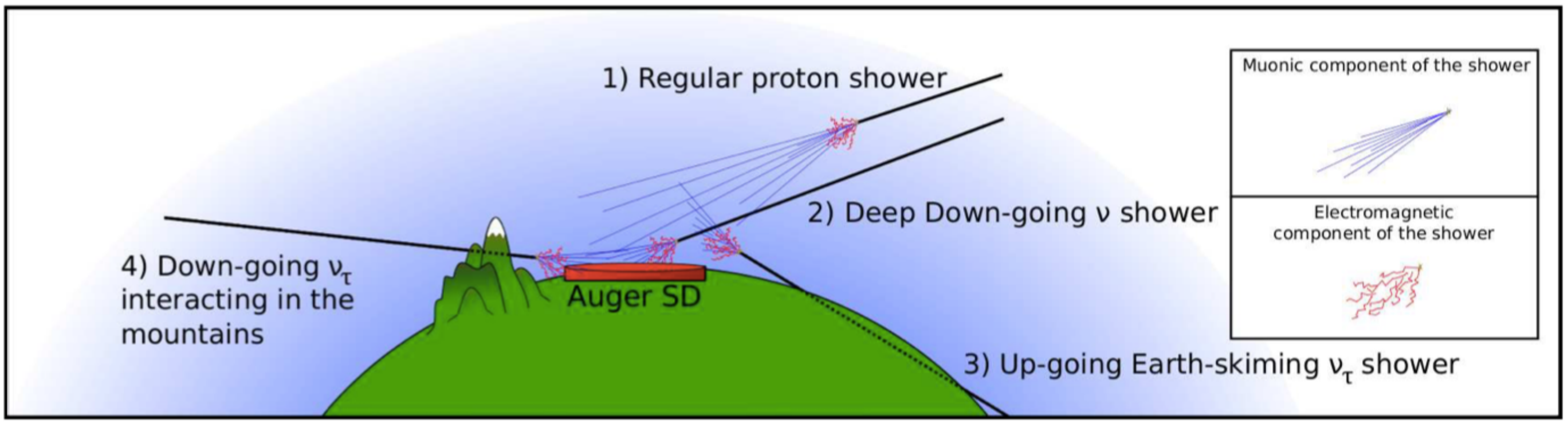}
\caption{Picture displaying the various neutrino detection modes (2, 3 and 4) in the Pierre Auger surface detector (SD, consisting of water Cherenkov tanks).}
\label{fig:paodetection}
\end{figure}

The PAO water Cherenkov tanks are sensitive to neutrino-induced air showers above $10^{17}$\,eV in two detection modes (see Figure \ref{fig:paodetection}):
\begin{itemize}
\item Downward-going neutrinos of all flavors interacting in the atmosphere for zenith angles of $75^{\circ}<\theta<90^{\circ}$, i.e. neutrinos travelling a relatively long distance in Earth's atmosphere.
\item Tau neutrino interactions in the Earth's crust that produce decaying tau leptons in the vicinity of the detector. This can happen either for so-called Earth-skimming events at zenith angles of $90^{\circ}<\theta<95^{\circ}$ or for downward-going tau neutrinos interacting in the mountains surrounding the detector.
\end{itemize}

The discrimination between cosmic ray induced air showers that originate from high up in the atmosphere and neutrino induced air showers that can decay anywhere in the atmosphere is maximised for nearly horizontal neutrinos that interact close to the detector. Those neutrino-induced air showers are in an early stage of development when detected and therefore still contain a relatively large electromagnetic component, giving a distinct timing signal in the water tanks. Instead, for the cosmic ray induced air showers, the electromagnetic component would have been absorbed higher up in the atmosphere already.

\begin{figure}[htb]
\centering
\includegraphics[width=0.8\textwidth, bb = 40 40 800 600]{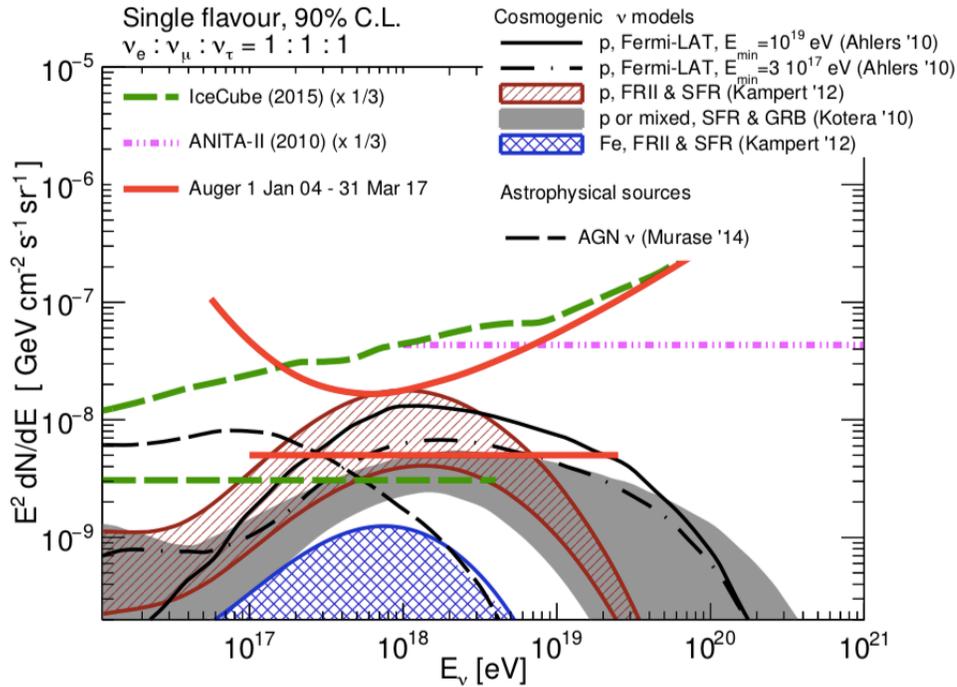}
\caption{90\% C.L. integral upper limits for a diffuse ultra-high energy neutrino flux $dN/dE_{\nu} = k E^{-2}$ (horizontal lines represent the normalization $k$) and differential upper limits (non-horizontal curves) for the Pierre Auger Observatory (red), ANITA-II (pink) and IceCube (green). Limits are quoted for a single flavor assuming equal flavor ratios. In addition, predictions for several neutrino models are shown. Plot source: \cite{paolimits}.}
\label{fig:paolimits}
\end{figure}

Analysing 13 years of PAO data, no ultra-high energy neutrino events have been detected. As a result, PAO has set neutrino flux upper limits at neutrino energies between $10^{17}$ and $10^{19}$\,eV \cite{paolimits}, see Figure \ref{fig:paolimits}. This neutrino energy range is complementary to that covered by IceCube and ANTARES. Updated extremely-high-energy neutrino flux limits from IceCube are available in \cite{icecubeehe}.

\section{Conclusion}
IceCube is at the forefront of neutrino astronomy since the discovery of an astrophysical neutrino flux in 2013 and first evidence for an astrophysical neutrino source in 2018. IceCube has found two tau neutrino candidates in their 7.5 year high-energy starting events dataset. One of these two candidates shows a striking double pulsed time signal in several optical modules surrounding the fitted double cascade. The fitted cosmic neutrino flavor ratio is $\nu_{e} : \nu_\mu : \nu_\tau = 0.29 : 0.50 : 0.29$, which within the uncertainties is consistent with the expected ratio of 1 :  1 :  1 for neutrino oscillations at cosmic baselines, but a zero $\nu_\tau$ flux can't be excluded either. At lower energies, IceCube is sensitive to tau neutrino appearance in atmospheric neutrino oscillations. The planned IceCube Upgrade will increase the sensitivity to the tau normalisation parameter and be able to measure the neutrino mass ordering.

The IceCube Upgrade science goals are competitive both in sensitivity and timeline with those of KM3NeT ORCA (atmospheric neutrino oscillations) and KM3NeT ARCA (astrophysical neutrinos), which are currently under construction at the bottom of the Mediterranean Sea. 

The Pierre Auger Observatory is sensitive to ultra-high energy ($>10^{17}$\,eV) cosmic neutrinos, in particular through the detection of Earth-skimming tau neutrinos, but so far no events have been detected yet.

\section*{Acknowledgements}
The author would like to thank the International Balzan Prize Foundation and Francis Halzen for their support.

\bibliography{Tau_BiBTeX_File.bib}

\nolinenumbers

\end{document}